\documentclass[twocolumn,preprintnumbers,amsmath,amssymb]{revtex4}
\usepackage{epsfig}
\usepackage{color}
\usepackage[colorlinks,urlcolor=blue,linkcolor=blue,citecolor=red]{hyperref}
\begin{document}
\setlength{\voffset}{1.0cm}
\title{Fermion number of twisted kinks in the NJL$_2$ model, revisited} 
\author{Michael Thies\footnote{michael.thies@gravity.fau.de}}
\affiliation{Institut f\"ur  Theoretische Physik, Universit\"at Erlangen-N\"urnberg, D-91058, Erlangen, Germany}
\date{\today}

\begin{abstract}
As a consequence of axial current conservation, fermions cannot be bound in localized lumps in the massless NJL$_2$ model. 
In the case of twisted kinks, this manifests itself
in a cancellation between valence fermion density and fermion density induced in the Dirac sea. To attribute
the correct fermion number to these bound states requires an infrared regularization. Recently, this has been 
achieved by introducing a bare fermion mass, at least in the non-relativistic regime of small twist angles and fermion
numbers. Here, we propose a simpler regularization by a finite box which preserves integrability and can be applied 
at any twist angle. A consistent and physically plausible assignment of fermion number to all twisted kinks emerges.
\end{abstract}
%\pacs{}
\maketitle

%<<<<<<<<<<<<<<<<<<<<<<<<<<<<<<<<<<<<<<<<<<<<<<<<<<<<<<<<<<<<<<<<<<<<<<<<<<<<<<<<<<<<<<<<<<<< <<<<<<<<<<<<<<<<<<<<<<<<<<<<<
%<<<<<<<<<<<<<<<<<<<<<<<<<<<<<<<<<<<<<<<<<<<<<<<<<<<<<<<<<<<<<<<<<<<<<<<<<<<<<<<<<<<<<<<<<<<<<<<<<<<<<<<<<<<<<<<<<<<<<<<<<<
\section{Why infrared problems?}
\label{sect1}
%<<<<<<<<<<<<<<<<<<<<<<<<<<<<<<<<<<<<<<<<<<<<<<<<<<<<<<<<<<<<<<<<<<<<<<<<<<<<<<<<<<<<<<<<<<<<<<<<<<<<<<<<<<<<<<<<<<<<<<<<<<
%<<<<<<<<<<<<<<<<<<<<<<<<<<<<<<<<<<<<<<<<<<<<<<<<<<<<<<<<<<<<<<<<<<<<<<<<<<<<<<<<<<<<<<<<<<<<<<<<<<<<<<<<<<<<<<<<<<<<<<<<<<

Consider the Nambu--Jona-Lasinio model \cite{1}  (NJL$_2$), or chiral Gross-Neveu model \cite{2} (GN), in 1+1 dimensions,
\begin{equation}
{\cal L}_{{\rm NJL}_2}  =  \bar{\psi}  i \partial \!\!\!/ \psi + \frac{g^2}{2}\left[  (\bar{\psi}\psi)^2 + (\bar{\psi} i \gamma_5 \psi)^2 \right] .
\label{1}
\end{equation}
Sums over $N$ flavors are suppressed as usual ($\bar{\psi}\psi = \sum_{k=1}^N \bar{\psi}_k \psi_k$ etc.). 
Since Lagrangian (\ref{1}) features only zero-range interactions, one would expect ultraviolet (UV) rather then 
infrared (IR) problems. Indeed, in 3+1 dimensions this theory is not renormalizable. In 1+1 dimensions, 
UV problems are harmless and can be handled by a mere renormalization of the coupling constant. In order to
understand the origin of possible IR problems, we have to remember that the NJL$_2$ model possesses a
continuous U(1) chiral symmetry,
\begin{eqnarray}
\psi & \to &  e^{i \alpha \gamma_5} \psi, \quad \gamma_5 = \gamma^0 \gamma^1,
\nonumber \\
\left( \bar{\psi}\psi - i \bar{\psi}i \gamma_5 \psi \right) & \to & e^{2i \alpha} \left( \bar{\psi}\psi - i \bar{\psi}i \gamma_5 \psi \right).
\label{2}
\end{eqnarray}
Strictly speaking, a continuous symmetry cannot be broken spontaneously in 1+1 dimensions \cite{3,4}. However, in the large $N$ limit,
mean field theory is believed to become exact so that one may envisage spontaneous breakdown of chiral symmetry
in the NJL$_2$ model \cite{5,6}. The appropriate mean field approach for fermions is the Hartree-Fock (HF) or time dependent Hartree-Fock (TDHF)
approximation. In the relativistic version needed here, one starts from the Dirac equation
\begin{equation}
\left( i  \partial \!\!\!/ - S -i \gamma_5 P \right) \psi = 0 .
\label{3}
\end{equation}
The scalar ($S$) and pseudoscalar ($P$) potentials obey the self-consistency conditions
\begin{eqnarray}
S & = &  - g^2 \langle \bar{\psi} \psi \rangle = - g^2 \sum_{\alpha}^{\rm occ} \bar{\psi}_{\alpha} \psi_{\alpha},
\nonumber \\
P & = &  - g^2 \langle \bar{\psi} i \gamma_5 \psi \rangle = - g^2 \sum_{\alpha}^{\rm occ} \bar{\psi}_{\alpha} i \gamma_5 \psi_{\alpha}.
\label{4}
\end{eqnarray}
The sum over all occupied orbits includes the Dirac sea and possible occupied positive energy levels. 
The vacuum corresponds to the solution of Eqs.~(\ref{3},\ref{4}) with homogeneous $S,P$. It is infinitely degenerate 
and characterized by a chiral vacuum angle $\theta$,
\begin{equation}
\Delta = S-iP = m e^{i \theta}.
\label{5}
\end{equation}
The U(1) manifold of all possible vacua is called the chiral circle. Its radius is the dynamical fermion mass $m$, generated
by dimensional transmutation from a dimensionless coupling constant via the vacuum gap equation \cite{2,7}
\begin{equation}
\frac{\pi}{Ng^2} = \ln \frac{\Lambda}{m}.
\label{6}
\end{equation}
Spontaneous symmetry breaking now amounts to picking one point on the chiral circle, say $\Delta=m$, as vacuum. 
Computing small fluctuations around the vacuum with the relativistic random phase approximation (RPA) 
yields the information about the meson spectrum. One finds a massive, scalar $\sigma$-meson ($m_{\sigma}=2 m$)
and a massless, pseudoscalar $\pi$-meson, the ``would-be" Goldstone boson \cite{8,9}. The latter corresponds to fluctuations
in the flat direction, here along the chiral circle. 

So far, everything is familiar from higher dimensions. A novel consequence of
chiral symmetry specific to low dimensions is the presence of massless baryons associated to a full turn around the 
chiral circle. If travelled ``infinitely slowly", this does not cost any energy \cite{7,10}. The easiest way to understand
these exotic objects is to work in a finite box of size $L$. Let us start from the vacuum HF solution and apply a local chiral transformation
[see Eq.~(\ref{2})] with linearly $x$-dependent phase $\alpha$,
\begin{equation}
\alpha = \frac{\pi}{L} x.
\label{7}
\end{equation}
As a consequence of the axial anomaly,
this yields a new, self-consistent HF solution with uniform fermion density
\begin{equation}
\frac{\rho}{N} = \frac{1}{\pi} \partial_x \alpha = \frac{1}{L}
\label{8}
\end{equation}
and hence a ``baryon" consisting of $N$ (strongly bound) fermions. This unusual 
``Goldstone baryon" is completely delocalized, with both mass
\begin{equation}
\frac{M_B}{N}= \frac{\pi}{2L}
\label{9}
\end{equation}
and fermion density (\ref{8}) vanishing in the limit $L \to \infty$. Nevertheless, it carries total fermion number $N$.
The fact that the fermions are delocalized is an unavoidable and well known consequence
of axial current conservation in 1+1 dimensions \cite{8,11}. In the massless NJL$_2$ model, we have the conservation laws
\begin{eqnarray}
\partial_{\mu} j_V^{\mu} & = & \partial_{\mu} \bar{\psi} \gamma^{\mu} \psi = 0,
\nonumber \\
\partial_{\mu} j_A^{\mu} & = & \partial_{\mu} \bar{\psi} \gamma^{\mu}\gamma_5 \psi = 0.
\label{10}
\end{eqnarray}
In 1+1 dimensions, vector and axial vector currents are related as follows, 
\begin{equation}
j_V^0 = j_A^1 := \rho, \quad j_V^1 = j_A^0 := j.
\label{11}
\end{equation}
Taking an expectation value of the 2nd line of Eq.~(\ref{10}) in a static configuration then implies that
\begin{equation}
\partial_x \langle \rho \rangle = 0.
\label{12}
\end{equation}
Hence fermions are always spread out over the whole volume.
While these massless baryons are specific for low dimensions,
they are not an artefact of the large $N$ limit, but have been identified in the NJL$_2$ model and even QCD$_2$ for finite $N$,
see Refs.~\cite{6,12}. The chiral phase $\alpha$ can be interpreted as a classical, macroscopic pion field. Winding number is baryon number, just as
in the Skyrme model \cite{13}. This whole construction has been useful mainly in the context of thermodynamics and finite density systems \cite{7,10,14},
where a periodic array of massless baryons gives rise to a ``chiral spiral"  type of mean field.
If one considers matter with a finite density of fermions, the relevant length scale for the size of a baryon is $1/\rho$ rather than $L$, so 
that there is no IR problem here.
The IR problem manifests itself as soon as we consider the limit $L \to \infty$ for a single baryon, Eqs.~(\ref{8},\ref{9}), where the density vanishes
but fermion number remains non-zero. It is simply impossible to ``see" the massless baryon if one works from the outset in the infinite volume limit. 
Nothing comparable happens in the standard GN model with discrete chiral symmetry, where there are no massless particles.

The other place where IR problems have been encountered is the ``twisted kink", originally found by Shei using inverse
scattering theory \cite{15}. This is a HF solution tracing out a chord between two arbitrary points on the chiral circle in the ($S,P$) plane.
By a proper choice of the global chiral angle, it can be cast into the form
\begin{equation}
\Delta = m \frac{e^{i\varphi} + e^{-i \varphi} e^{2\xi}}{1 + e^{2\xi}} = m \left( \cos \varphi - i \sin \varphi \tanh \xi \right)
\label{13}
\end{equation}
with
{\begin{equation}
\xi = m x \sin \varphi
\label{14}
\end{equation}
in the rest frame. The potential $\Delta$ interpolates between two vacua at $\theta= \varphi$ ($x \to - \infty$) and $\theta = - \varphi$
($x \to \infty$). The kink potential has a single bound state which can be filled with $N_0\in [0,N]$ fermions. Self-consistency
relates the filling fraction $\nu=N_0/N$ (a continuous parameter in the large $N$ limit) to the ``twist angle" $\varphi$ 
\begin{equation}
\nu = \frac{\varphi}{\pi}.
\label{15}
\end{equation}
If one evaluates the fermion density, one finds that the valence density from the bound state  
is cancelled exactly by the density induced in the Dirac sea (arising from the negative energy continuum states) \cite{11,16}.
This led to the conclusion that the twisted kink does not carry fermion number and hence should not be regarded as a baryon.

Recently, an attempt was made to extend the twisted kinks to finite bare fermion masses \cite{17}. Due to technical difficulties,
this could only be done in the non-relativistic limit so far, i.e., for small occupation fraction and
smooth shapes. The vacuum being unique in the massive model no matter how small the bare mass is, this must
be accompanied by an untwisting of the kink. In the limit where the bare mass vanishes, this can also be interpreted 
as a kind of IR regularization. The new insight was 
that fermion density of the massive, untwisted kink is spread out over a pion cloud of size $1/m_{\pi}$. If one lets 
$m_{\pi} \to 0$, the density vanishes but the volume diverges, fermion number being kept at a non-zero value.
As a result, twisted kinks carry the same fermion number as what was put into the valence level. In the naive 
chiral limit, one only sees the screening of the valence charge but misses the spread out charge distribution, just like in the 
case of the massless baryon. Although the calculations could only be done for small occupation fraction, this led to the 
conjecture that the correct picture of baryons in the chiral limit of the NJL$_2$ model is a composite
of a chord soliton \`a la Shei and a fractional turn of the chiral spiral. In the present paper, we
will show how to realize this picture quantitatively for arbitrary occupation numbers. We shall use a finite
box IR regularization as for the massless baryon. Thereby, integrability of the model is preserved and computations are much easier than
solving the massive NJL$_2$ model.
 
%<<<<<<<<<<<<<<<<<<<<<<<<<<<<<<<<<<<<<<<<<<<<<<<<<<<<<<<<<<<<<<<<<<<<<<<<<<<<<<<<<<<<<<<<<<<< <<<<<<<<<<<<<<<<<<<<<<<<<<<<<
%<<<<<<<<<<<<<<<<<<<<<<<<<<<<<<<<<<<<<<<<<<<<<<<<<<<<<<<<<<<<<<<<<<<<<<<<<<<<<<<<<<<<<<<<<<<<<<<<<<<<<<<<<<<<<<<<<<<<<<<<<<
\section{A simple infrared regularization}
\label{sect2}
%<<<<<<<<<<<<<<<<<<<<<<<<<<<<<<<<<<<<<<<<<<<<<<<<<<<<<<<<<<<<<<<<<<<<<<<<<<<<<<<<<<<<<<<<<<<<<<<<<<<<<<<<<<<<<<<<<<<<<<<<<<
%<<<<<<<<<<<<<<<<<<<<<<<<<<<<<<<<<<<<<<<<<<<<<<<<<<<<<<<<<<<<<<<<<<<<<<<<<<<<<<<<<<<<<<<<<<<<<<<<<<<<<<<<<<<<<<<<<<<<<<<<<<

A proven remedy against IR problems is a finite volume. In one space dimension, consider an interval of length $L$
on the $x$-axis, say $x \in [-L/2,L/2]$. We then need to specify boundary conditions for the fermion fields
of the NJL$_2$ model. If one is interested in twisted kinks, one might be tempted to impose chirally twisted 
boundary conditions,
\begin{equation}
\psi(x=L/2,t) = e^{i \theta \gamma_5} \psi(x=-L/2,t).
\label{16}
\end{equation}
This would imply that the mean field obeys
\begin{equation}
\Delta(L/2,t) = e^{2i \theta} \Delta(-L/2,t). 
\label{17}
\end{equation}
Thus only one particular value of the twist would be allowed. This would be appropriate in situations where the 
asymptotic vacuum angles are prescribed externally, e.g., a superconducting wire connecting two
bulk superconductors with different macroscopic phases. From an intrinsic field theory point of view, it is problematic.
First of all, the standard vacuum ($\Delta=m$) cannot be accommodated in the same space as the twisted kink. This 
makes it impossible to evaluate quantities which need a careful counting of modes and vacuum subtraction,
like masses or fermion numbers.
Besides, one would like to describe different twisted kinks and their interactions in a common space, otherwise they
would belong to disjoint theories. This incites us to go back to periodic (or quasi-periodic) boundary conditions for spinors. But then $\Delta$ 
has to be periodic. We are back at the same situation as in the case of finite bare mass, namely we have to 
use a unique vacuum asymptotically. Twist can then only survive in the sense of intermediate asymptotics, as was found in Ref.~\cite{17}.
We propose to use once again the trick with the local chiral rotation as follows:
Suppose we have found a HF solution in the infinite volume with total twist angle $\varphi$,
\begin{equation}
\lim_{x \to \infty} \frac{\Delta(x,t)}{\Delta(-x,t)} = e^{-2i \varphi}.
\label{18}
\end{equation}
It could comprise one or several twisted kinks and breathers and be either static or time dependent, but we do not 
need to specify any details at this point.
Introduce a length $L$ such that the limit (\ref{18}) is reached for any practical purpose at $x=L/2$. 
Then perform a chiral rotation which undoes the twist in the smoothest possible way, namely with a linear $x$ dependence,
\begin{eqnarray}
\psi(x,t) & \to & \tilde{\psi}(x,t) =  e^{i \varphi x/L \gamma^5} \psi(x,t),
\nonumber \\
\Delta(x,t) & \to & \tilde{\Delta}(x,t) = e^{2i \varphi x/L} \Delta(x,t).
\label{19}
\end{eqnarray}
The new mean field $\tilde{\Delta}$ is now periodic by construction,
\begin{equation}
\tilde{\Delta}(L/2,t)=\tilde{\Delta}(-L/2,t).
\label{20}
\end{equation}
If one is really interested in a finite box, one has to discretize momenta in accordance with the boundary conditions chosen.
Here we use $L$ only as a regulator and are ready to choose arbitrarily large values of $L$. This should enable us to 
take over the wave functions $\psi(x,t)$ and mean fields $\Delta(x,t)$ 
from a standard continuum calculation. Due to the axial anomaly however, the transformation (\ref{19}) generates a 
constant fermion density 
\begin{equation}
\frac{\rho}{N} = \frac{1}{\pi} \partial_x \alpha = \frac{\varphi}{\pi} \frac{1}{L}  
\label{21}
\end{equation}
and thus the fermion number 
\begin{equation}
\frac{N_f}{N} = \int_{-L/2}^{L/2} dx \frac{\rho}{N} = \frac{\varphi}{\pi}.
\label{22}
\end{equation}
Equivalently, a local chiral transformation with linear $x$-dependence amounts to introducing a chemical potential $\mu= \pi \rho$
into the system.

The main advantage of the proposed IR regularization as compared to introducing a bare fermion mass 
is its simplicity. We can generate an exact HF solution in this way, free of IR ambiguities, without loosing
integrability. The HF spinors and the mean field without the regulator can be taken over from the naive continuum
calculation. All one has to do is apply the transformation (\ref{19}) and take the limit $L\to \infty$ at
the end of the calculation. Twist does not disappear, but is still present as intermediate asymptotics.
The constant fermion density is consistent with chiral symmetry and axial current conservation.
The implication for the physics interpretation is nevertheless far reaching: All states with a twist
in the naive continuum limit acquire a constant, inert fermion density $\sim 1/L$. These fermions
are just spectators to what is going on, but are important for the classification of states and for avoiding
certain paradoxes. Most importantly, this opens the way to describing baryonic states with fermion
numbers $N_f<N$. They are expected to appear in a non-confining theory like the NJL$_2$ model
and have been known in the GN model with discrete chiral symmetry since the work of Dashen,
Hasslacher and Neveu \cite{18}. In QCD$_2$, due to confinement, such objects
cannot exist as they would not be color singlets, at least if one interprets $N$ as the number of colors.

In the following section, we will show in more detail how this regularization works in the case of a single
twisted kink.
 
%<<<<<<<<<<<<<<<<<<<<<<<<<<<<<<<<<<<<<<<<<<<<<<<<<<<<<<<<<<<<<<<<<<<<<<<<<<<<<<<<<<<<<<<<<<<< <<<<<<<<<<<<<<<<<<<<<<<<<<<<<
%<<<<<<<<<<<<<<<<<<<<<<<<<<<<<<<<<<<<<<<<<<<<<<<<<<<<<<<<<<<<<<<<<<<<<<<<<<<<<<<<<<<<<<<<<<<<<<<<<<<<<<<<<<<<<<<<<<<<<<<<<<
\section{Application to the twisted kink}
\label{sect3}
%<<<<<<<<<<<<<<<<<<<<<<<<<<<<<<<<<<<<<<<<<<<<<<<<<<<<<<<<<<<<<<<<<<<<<<<<<<<<<<<<<<<<<<<<<<<<<<<<<<<<<<<<<<<<<<<<<<<<<<<<<<
%<<<<<<<<<<<<<<<<<<<<<<<<<<<<<<<<<<<<<<<<<<<<<<<<<<<<<<<<<<<<<<<<<<<<<<<<<<<<<<<<<<<<<<<<<<<<<<<<<<<<<<<<<<<<<<<<<<<<<<<<<<

As basic example for the IR regularization, consider a single twisted kink at rest. The original form of the 
mean field $\Delta=S-iP$ has been given in (\ref{13}). According to the prescription proposed in Sect.~\ref{sect2}, it
should be replaced by the IR regularized expression 
\begin{equation}
\tilde{\Delta} = m e^{2i\varphi x/L} \frac{e^{i\varphi} + e^{-i \varphi} e^{2\xi}}{1+ e^{2 \xi}}
\label{23}
\end{equation}
with $\xi$ from Eq.~(\ref{14}).
%###########################################################################################################################
\begin{figure}
\begin{center}
\epsfig{file=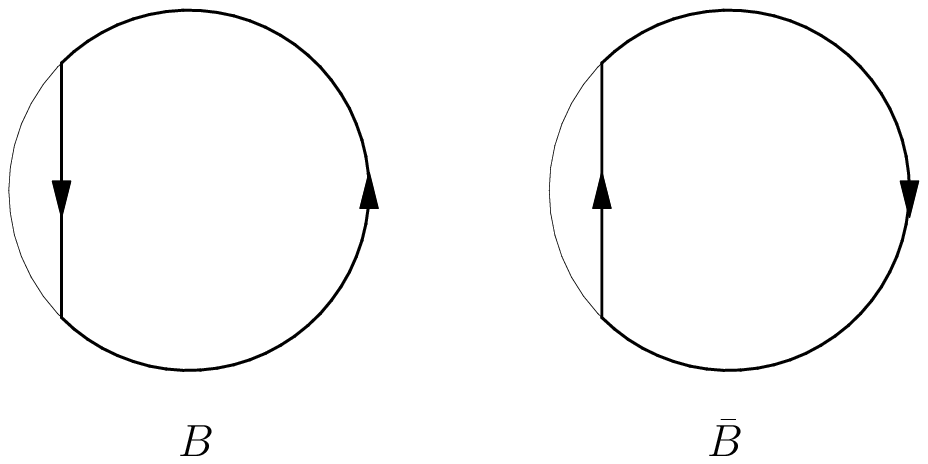,width=8cm,angle=0}
\caption{($S,P$)-trajectory of IR regulated twisted kink. The chord is the standard twisted kink, the arc  
an incomplete turn of the chiral spiral arising from the regularization. Baryon (left) and antibaryon (right) differ
through the orientation of the loops. Parameters: $\varphi=3\pi/4, L\to \infty$.}
\label{fig1}
\end{center}
\end{figure}
%############################################################################################################################
If we now plot the ($S,P$)-trajectory, we find that the regularization does the same thing as a small bare
mass --- the chord corresponding to the original twisted kink is closed by an arc along the chiral circle. This is shown in Fig.~\ref{fig1} for $\varphi=3\pi/4$.
Baryon and antibaryon differ only in the orientation of the loop ($x\to -x$). Unlike the results for finite fermion mass in Ref.~\cite{17}, this construction applies
to any twist angle $\varphi \in [0,\pi]$. 

Let us check more quantitatively the picture that a NJL$_2$ baryon is a twisted kink glued to an incomplete
winding of the chiral spiral. To this end, we use the dimensionless variable $\xi$ of Eq.~(\ref{14}) as spatial parameter, rather than $x$, 
\begin{equation}
\tilde{\Delta} = m \exp{\left\{\frac{2i\varphi \xi}{m L \sin \varphi}\right\}} \frac{e^{i\varphi} + e^{-i \varphi} e^{2\xi}}{1+ e^{2 \xi}}.
\label{24}
\end{equation}
We divide the finite box ($x\in [-L/2,L/2]$) into two regions. In the inner region ($|\xi|< \xi_0$), we require that the difference between
the original twisted kink and its spatial asymptotics differs by less than $\epsilon$. This is fulfilled if we choose
\begin{equation}
\xi_0 = m x_0 \sin \varphi =  - \frac{1}{2} \ln \frac{\epsilon}{2}.
\label{25}
\end{equation}
Besides, we demand that the correction from the IR regulator, Eq.~(\ref{19}), at the ``matching radius" $\xi_0$
be smaller than $\eta$. This gives a lower bound for $L$,
\begin{equation}
L \ge \frac{2 \varphi \xi_0}{\eta m \sin \varphi}.
\label{26}
\end{equation}
For these values of $\xi_0,L$, we may use the following patched approximation to (\ref{24}),
\begin{equation}
\tilde{\Delta} = \left\{  \begin{array}{cc}   m   e^{i \varphi}  e^{2i\varphi x/L} & ( -L/2 < x < - x_0) \\
m  \frac{e^{i\varphi} + e^{-i \varphi} e^{2\xi}}{1+ e^{2 \xi}} &  ( -\xi_0 < \xi < \xi_0) \\
m e^{-i \varphi} e^{2i\varphi x/L}  &  ( x_0 < x  < L/2) \end{array} \right.
\label{27}
\end{equation}
By way of example, choose $\epsilon = \eta = 10^{-4}, \varphi=\pi/2, m=1$. This yields $\xi_0 = 4.95, L \ge 1.56 \times 10^5$.
The piecewise approximation (\ref{27}) is indistinguishable on a plot from the full, regularized expression (\ref{23}), even
in the region around $\xi_0$ where the patching occurs.  
For $\varphi$ approaching the maximum value of $\pi$, $L$ blows up due to the $1/\sin \varphi$ factor in (\ref{26}).
Since we are only interested in the limit $L \to \infty$, the accuracy can be arbitrarily improved
and the picture of a twisted kink chord soliton glued to a fractional turn of the circular chiral spiral
proposed in Ref.~\cite{17} is fully confirmed.

If one decreases $L$ continuously, the corners between chord and arc get rounded off, see Fig.~\ref{fig2}.
This reminds us qualitatively of the findings  for decreasing bare fermion masses in Ref.~\cite{17}.
IR regularization via bare fermion mass and finite box are thus consistent, giving the same result
in the limit $m_0 \to 0$ and $L\to \infty$, respectively.  

%###########################################################################################################################
\begin{figure}
\begin{center}
\epsfig{file=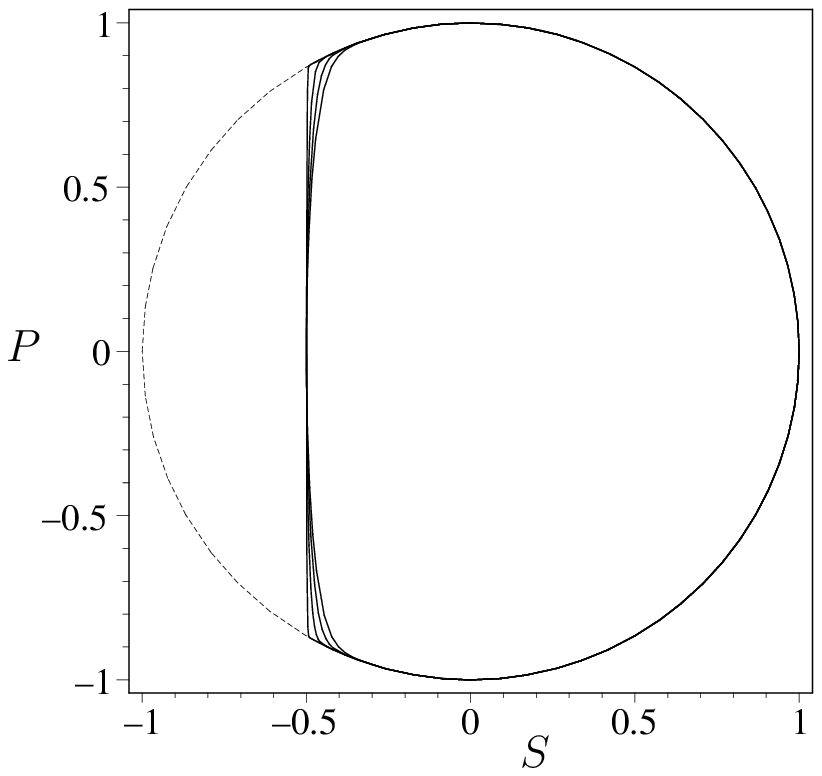,width=6cm,angle=0}
\caption{Dependence of the twisted kink contour on the IR regulator $L$. Parameters: $m=1, \varphi=2\pi/3, L=100,150,300,2000$.
The smaller $L$, the smoother the transition region between chord and arc.}
\label{fig2}
\end{center}
\end{figure}
%############################################################################################################################

%###########################################################################################################################
\begin{figure}
\begin{center}
\epsfig{file=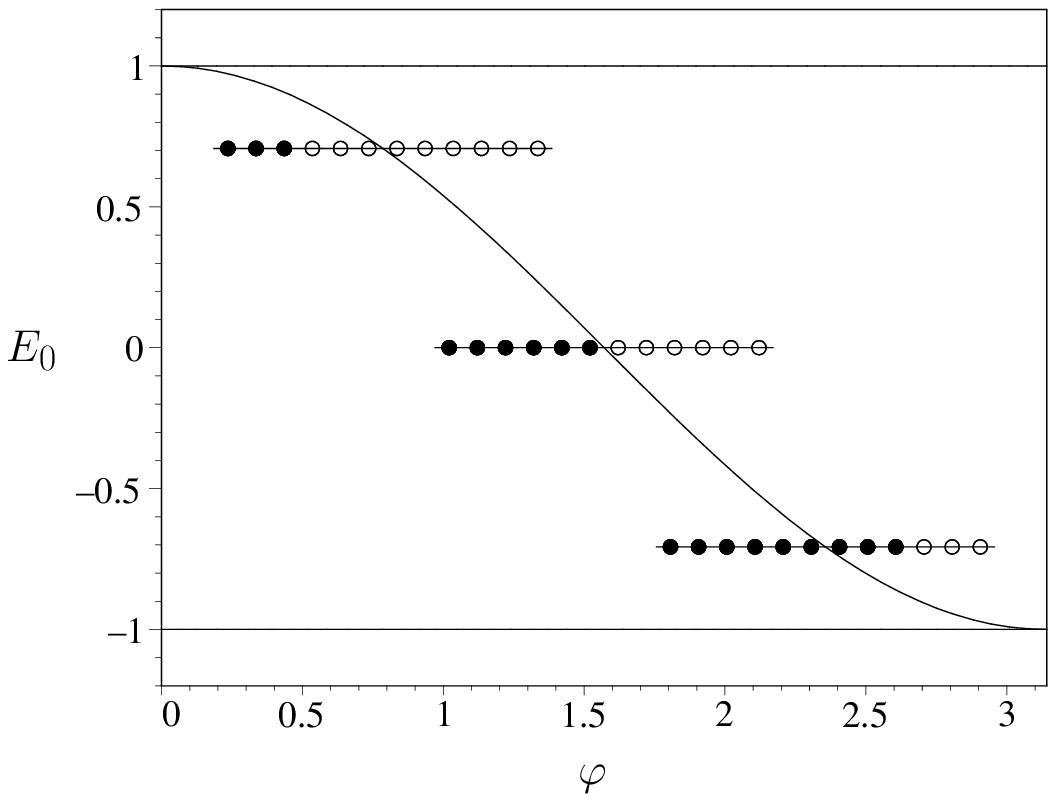,width=8cm,angle=0}
\caption{Bound state energy $m\cos \varphi$ of twisted kink versus twist angle, in units where $m=1$.
The three levels shown belong to $N=12, \varphi= \pi/4, \pi/2, 3\pi/4$ and illustrate the occupation
with valence fermions. For this value of $N$, $\varphi$ is discretized in steps of $\pi/12$
and the occupation number increases from 0 ($E_0=m$) to 12 ($E_0=-m$) in integer steps.}
\label{fig3}
\end{center}
\end{figure}
%############################################################################################################################

The main discrepancy between the ``naive" twisted kink and the IR regularized version concerns fermion number.
It may be worthwhile to address this issue in more detail, using the single twisted kink as example. 
To set the stage for the following discussion, we briefly recall the history of the twisted kink and
the fermion number attributed to it.
We begin with Shei's original point of view \cite{15}. This author ignores fermion number induced in the Dirac sea
by the twisted kink potential and looks only at the valence fermion density.
As one increases $\varphi$ from 0 to $\pi$, the bound state energy $E_0=m \cos \varphi$ crosses the gap,
with occupation fraction $\nu=\varphi/\pi$, see Fig.~\ref{fig3}. At $\varphi=\pi/2$, the energy $E_0$ crosses 0. Shei counts the
number of valence fermions until this point is reached, associating fermion number $N\varphi/\pi$
with the twisted kink for $\varphi \in [0,\pi/2]$. The maximum fermion number of $N/2$ is reached at mid-gap, half filling ($\varphi=\pi/2$). For negative values of $E_0$
($\varphi>\pi/2$), he counts the number of ``holes", assigning fermion number $N(\varphi/\pi-1)$ ranging from $-N/2$ to 0 as $\varphi$ goes from $\pi/2$ to $\pi$.
This way of reasoning would be perfectly correct in the GN model where there is no induced fermion density,
but cannot be justified in the NJL$_2$ model.
Subsequently, sea effects were taken into account by other authors and an exact cancellation between valence
fermion density and induced density was found \cite{11,16}. Since axial and vector current conservation admit only
spatially constant fermion density, it was concluded that the density vanishes identically --- twisted kinks
seemed to carry no fermions at all. They were interpreted as ``baryonium states" rather than baryons.
From the physics point of view, both of these scenarios are somewhat puzzling: If one adds
1,2,...,$N$ fermions to the vacuum, one winds up with the vacuum --- where have all the fermions gone?
Actually, Fig.~\ref{fig3} already gives a clue to the answer. At $\varphi=\pi$, the bound state level is fully occupied ($N$
fermions) and touches the surface of the Dirac sea from above ($E_0=-m$). But this is exactly the picture expected from the
massless baryon, where a full turn around the chiral circle heaves up one occupied level from the sea 
by the axial anomaly. 

We now understand that a vanishing density does not necessarily imply vanishing fermion number, in the limit $L \to \infty$.
In the IR regularized version, fermions are spread out over the whole volume, but their total number agrees with the 
number of valence fermions ($1...N$ for $\varphi=0...\pi$).  Antibaryons can be generated by changing the 
orientation of the loop in the ($S,P$)-plane ($x \to - x$) for the same values of $\varphi$. The fact that kinks carrying $N_f$ and $N-N_f$ fermions are
degenerate follows from the observation that the mass depends only on the length of the chord, the
fermion number on the length of the arc. 

As a matter of fact, we can extend the range of allowed fermion numbers as follows. Multiplying $\tilde{\Delta}$ by the phase factor 
\begin{equation}
p=e^{2i \pi x/L}
\label{28} 
\end{equation}
results in a new HF solution with an additional massless baryon. As long as the number of baryons is not
macroscopic, we can generate states with larger fermion number (by a multiple of $N$) simply by multiplying $\Delta$ by 
the $n$-th power of $p$. This corresponds to a chiral spiral with $n$ turns in the whole volume.
Such a construction only works because the massless baryons are non-interacting. 

Consider next an antibaryon with fermion number $- N \varphi/\pi$
(Eq.~(\ref{23}) with $x \to -x$ so as to reverse the orientation), 
\begin{eqnarray}
\tilde{\Delta} & = & m e^{-2i\varphi x/L} \frac{e^{i\varphi} + e^{-i \varphi} e^{-2\xi}}{1+ e^{-2 \xi}}
\nonumber \\
& = &  m e^{-2i\varphi x/L} \frac{e^{-i\varphi} + e^{i \varphi} e^{2\xi}}{1+ e^{2 \xi}}.
\label{29}
\end{eqnarray}
It has the same mass as the baryon with twist angle $\varphi$, but opposite fermion number.
Again, one can extend it to larger negative fermion numbers by adding massless antibaryons.

%###########################################################################################################################
\begin{figure}
\begin{center}
\epsfig{file=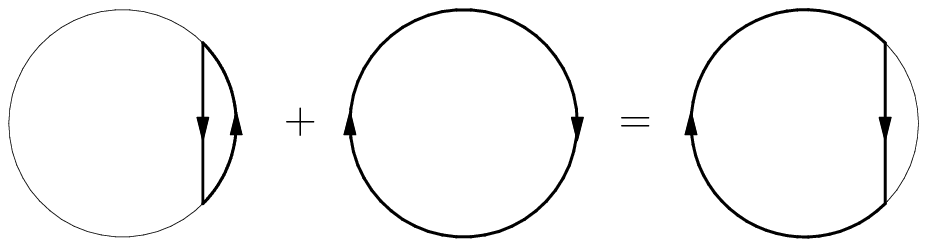,width=8cm,angle=0}
\caption{Graphical illustration of the degeneracy between twisted kinks with twist angles $\varphi$ and 
$\pi - \varphi$, here for $\varphi=\pi/4$. The full circle is a massless, non-interacting baryon. See main text
for a discussion}
\label{fig4}
\end{center}
\end{figure}
%############################################################################################################################

Finally, let us try to understand the degeneracy between kinks with twist angles $\varphi$ and $\pi-\varphi$ in different terms.
Start from the regularized $\varphi$-kink and add a massless antibaryon (Fig.~\ref{fig4})
\begin{equation}
\tilde{\Delta} = m  e^{-2i\pi x/L}  e^{2i\varphi x/L}   \frac{e^{i\varphi} + e^{-i \varphi} e^{2\xi}}{1+ e^{2 \xi}}.
\label{30}
\end{equation}
This differs from the mean field of an antibaryon with twist angle $\pi - \varphi$ by a discrete chiral transformation 
($\psi \to \gamma_5 \psi, \tilde{\Delta} \to - \tilde{\Delta}$), corresponding to a reflection at the origin in the figure.  Such an antibaryon
has the same mass as the 
baryon with twist $\pi - \varphi$. This does not give any new result, but underlines that the regulator factor
has been chosen consistently. With similar graphical arguments, one can interpret an antibaryon
with twist angle $-\varphi$ as a composite of a full antibaryon ($-\pi$) and a baryon with twist ($\pi-\varphi$). 
It is amusing that this takes us almost back to Shei's original identification of the $(\pi- \varphi)$-kink with the 
$(-\varphi)$-antikink. Indeed, these two states only differ by the addition of a non-interacting, massless antibaryon.

%<<<<<<<<<<<<<<<<<<<<<<<<<<<<<<<<<<<<<<<<<<<<<<<<<<<<<<<<<<<<<<<<<<<<<<<<<<<<<<<<<<<<<<<<<<<< <<<<<<<<<<<<<<<<<<<<<<<<<<<<<
%<<<<<<<<<<<<<<<<<<<<<<<<<<<<<<<<<<<<<<<<<<<<<<<<<<<<<<<<<<<<<<<<<<<<<<<<<<<<<<<<<<<<<<<<<<<<<<<<<<<<<<<<<<<<<<<<<<<<<<<<<<
\section{Summary and conclusions}
\label{sect4}
%<<<<<<<<<<<<<<<<<<<<<<<<<<<<<<<<<<<<<<<<<<<<<<<<<<<<<<<<<<<<<<<<<<<<<<<<<<<<<<<<<<<<<<<<<<<<<<<<<<<<<<<<<<<<<<<<<<<<<<<<<<
%<<<<<<<<<<<<<<<<<<<<<<<<<<<<<<<<<<<<<<<<<<<<<<<<<<<<<<<<<<<<<<<<<<<<<<<<<<<<<<<<<<<<<<<<<<<<<<<<<<<<<<<<<<<<<<<<<<<<<<<<<<

In this paper, we have proposed a simple IR regularization of twisted kinks in the massless NJL$_2$ model.
The main new result is the assignment of fermion number to twisted kinks. By promoting these objects to baryons carrying
$N_f < N$ fermions, we bridge the gap between two apparently unrelated constructs in the literature: Shei's twisted
kink, so far believed to have vanishing fermion number, and the axial anomaly driven, chiral spiral type of baryon carrying the maximum of $N$ fermions.
To simplify the language, let us call $N_f/N$ ``baryon number". Then baryons with integer baryon number are massless, non-interacting 
and delocalized objects, described by a full turn around the chiral circle in the ($S,P$) plane. Baryons with fractional baryon number
$\nu$ can now be identified with IR regularized twisted kinks with the twist angle $\varphi=\nu \pi$. They are massive and interacting, although they do not
interact with the massless, ``integer" baryons. In the ($S,P$) plane, they consist of a partial turn around the chiral circle, the end points being 
connected by a standard twisted kink, a chord soliton. This answers two questions at the same time: What determines the twist angle of twisted 
kinks, and what is the ground state of $N_f<N$ fermions in the NJL$_2$ model? Both of these questions could not be answered so far.
Now we see that the twist angle is simply determined by the total fermion number (an observable), which happens to coincide with the number of valence fermions.
The (IR regulated) twisted kink is nothing but the baryon with fractional baryon number.

As we have seen, twisted kinks with larger twist angles can be decomposed into a twisted kink with a smaller twist and one or several massless,
``integer" baryons. If one interprets these states as multi-baryon states, the single baryon or antibaryon states with fractional baryon number 
are those with $\varphi \in [-\pi/2,\pi/2]$. This is a kind of ``fundamental domain" in twist space. All other twists differ by an integer multiple of $\pi$
and therefore by addition of massless (anti-)baryons. The mass $m \cos \varphi$ is the same for $\pm \varphi$, a fact that can now be related to 
baryon/antibaryon symmetry. Superficially, this agrees with Shei's original claim, but we hope to have shown that things are more
subtle than previously thought.

In recent years, a lot of effort has been devoted to studying interactions of several twisted kinks, both in condensed matter \cite{19} and particle physics \cite{20,21}.
Because the fermions we have been discussing here are just a background of inert ``spectators" to whatever happens to the mean field,
we do not expect any significant impact of the present findings on these results.

%#################################################################################################################################

%#################################################################################################################################

\end{document}